\documentclass{aa}  
\bibpunct{(}{)}{;}{a}{}{,}
\usepackage[singlelinecheck=off]{caption}
\usepackage{graphicx}
\usepackage{longtable}
\usepackage{leftidx}
\usepackage{amssymb}
\usepackage{amsmath}
\usepackage{gensymb}
\usepackage{textcomp}
\usepackage[usenames,dvipsnames]{xcolor}
\usepackage{epstopdf}

\usepackage{txfonts}
\usepackage{mhchem}

\begin{document}

   \title{A sequential acid-base (SAB) mechanism in the interstellar medium: The emergence of cis formic acid in dark molecular clouds}

   \author{J. García de la Concepción
          \inst{1}
          \and
          I. Jiménez-Serra
          \inst{1}
          \and
          J. C. Corchado
          \inst{2}
          \and
          G. Molpeceres
          \inst{3}
          \and
          A. Martínez-Henares
          \inst{1}
          \and
          V. M. Rivilla
          \inst{1}
          \and
          L. Colzi
          \inst{1}
          \and
          J. Martín-Pintado
          \inst{1}
          }

   \institute{Centro de Astrobiología (CSIC-INTA), Ctra. de    Ajalvir Km. 4, Torrejón de Ardoz, 28850 Madrid, Spain\\
              \email{jgarcia@cab.inta-csic.es}
              \and
        Departamento de Ingeniería Química y Química Física, Facultad de Ciencias, and ICCAEx, Universidad Extremadura, Badajoz, Spain
         \and
        Institute for Theoretical Chemistry, University of Stuttgart, Pfaffenwaldring 55, 70569 Stuttgart, Germany
         }

   \date{}

 
  \abstract
   {The different abundance ratios between isomers of an organic molecule observed in the interstellar medium (ISM) provides valuable information about the chemistry and physics of the gas and eventually, the history of molecular clouds. In this context, the origin of an abundance of cis formic acid (c-HCOOH) of only 6\% the trans isomer (t-HCOOH) abundance in cold cores, remains unknown.}
   {In this work, we explain the presence of c-HCOOH in dark molecular clouds through the destruction and back formation of c-HCOOH and t-HCOOH in a cyclic process that involves HCOOH and highly abundant molecules such as HCO$^{+}$ and NH$_{3}$.}
   {We use high-level $ab$ $initio$ methods to compute the potential energy profiles for the cyclic destruction/formation routes of c-HCOOH and t-HCOOH. Accurate global rate constants and branching ratios were calculated based on the transition state theory and the master equation formalism under the typical conditions of the ISM.}
   {The destruction of HCOOH by reaction with HCO$^{+}$ in the gas phase leads to three isomers of the cation HC(OH)$_{2}^{+}$. The most abundant cation can react in a second step with other abundant molecules of the ISM like NH$_{3}$ to form back c-HCOOH and t-HCOOH. This mechanism explains the formation of c-HCOOH in dark molecular clouds. Considering this mechanism, the fraction of c-HCOOH with respect t-HCOOH is 25.7\%. To explain the 6\% reported by the observations we propose that further destruction mechanisms of the cations of HCOOH by collisions with abundant molecules or interconversion reactions on dust grains should be taken into account.}
   {The sequential acid-base (SAB) mechanism proposed in this work involves fast processes with very abundant molecules in the ISM. Thus, HCOOH very likely suffers our proposed transformations in the conditions of dark molecular clouds such as B5 and L483. This is a new approach in the framework of the isomerism of organic molecules in the ISM which has the potential to try to explain the ratio between isomers of organic molecules detected in the ISM.}

                
\titlerunning{A sequential acid-base (SAB) mechanism in the ISM}
    \maketitle

\section{Introduction}

The reaction mechanisms associated to the isomerism of organic molecules in the interstellar medium (ISM) are gaining the attention of the astrophysical community. Understanding the chemical processes that control the ratios found for isomers in the ISM gives valuable information for the chemistry and physical processes taking place in molecular clouds. To explain the isomer ratios observed in the ISM of some organic molecules, several processes have been invoked such as isomerization processes \citep{juan2021a,juan2021b,Molpeceres2021}, different formation routes for both isomers \citep{balucani18,baiano20,lupi20,Molpeceres2021}, or higher destruction rates for one of the isomers  \citep{Shingledecker2019,Shingledecker2020,Loomis2015}.
Recent quantum chemical studies have revealed that the E/Z and cis/trans ratio of the imines and some acids found in the ISM should be due to intramolecular isomerization reactions \citep{juan2021a,juan2021b}. These processes allow to reach the chemical equilibrium in the ISM between isomers for temperature ranges of $T_{\rm kin}$\(\sim\)200-400 K.  However, this process does not explain why the cis isomer of formic acid (c-HCOOH) is found to be only  \(\sim\)6\% of the trans isomer (t-HCOOH) abundance in cold and dense molecular clouds like Barnard 5 \citep[B5 hereafter;][]{taquet17} and L483 \citep{agundez19}, where the gas kinetic temperature is very low \citep[$T_{\rm kin}$ \(\sim\)5-15 K;][]{juan2021b}. 

Formic acid presents two isomers, trans (t-HCOOH) and cis (c-HCOOH), depending on the position of the hydrogen of the OH bond. The trans isomer (t-HCOOH) is the most stable one by 4.04 kcal mol$^{-1}$ (2033 K) \citep{juan2021b}. According to the equilibrium constant ($K_{eq}$) found between the two isomers at very low temperatures \citep[$T_{\rm kin}$ \(\sim\)5-15 K;][]{juan2021b}, the equilibrium should be completely displaced towards t-HCOOH ($K_{eq}$(10 K)=2.09x10$^{88}$). Indicating that the temperature is too low for allowing intramolecular isomerizations \citep{juan2021b}.

t-HCOOH has been widely detected in the ISM (see \citealp[]{juan2021b} and references therein). In contrast, c-HCOOH has been only detected in the photon-dominated region (PDR) of the Orion Bar \citep{cuadrado16} and in the cold dark clouds L483 \citep{agundez19} and B5 \citep{taquet17}. The abundance of c-HCOOH with respect to t-HCOOH in the  Orion Bar PDR is \(\sim\)33\%. In this case, the presence of c-HCOOH in the gas phase has been explained by photo-switching mechanisms because of the higher prevalence of UV photons \citep{cuadrado16}.

Most of the chemistry in cold sources such as B5 and L483 is dominated by gas phase chemistry. However, the formation of HCOOH \citep{Ioppolo2010,taquet17} and its sulphur derivative (HCOSH) \citep{Nguyen2021,Molpeceres2021} occur efficiently on dust grains. Although it has not been possible to  experimentally constrain what isomer of HCOOH predominates in its formation. \citet{Molpeceres2022} has recently demonstrated that c-HCOOH is destroyed on grain surfaces by reaction with hydrogen atoms. Thus, there should be a notable enrichment of the t-HCOOH isomer on dust grains. This finding indicates that the relative abundance between c-HCOOH and t-HCOOH cannot be explained through these grain surface destruction mechanism. Even so, HCOOH should not withstand the conditions found on dust grains \citep{Molpeceres2022}. Therefore, it is unclear how c-HCOOH is present in the gas phase toward these very cold sources.

It has been suggested that a plausible route of gas phase formation of HCOOH in dark molecular clouds is through the dissociative recombination of the OH-protonated formic acid (HCOOH$_{2}^{+}$) \citep{vigren2010}. However, the latter study does not discriminate between the isomers. Furthermore, the  dissociative recombination of  HCOOH$_{2}^{+}$ gives HCOOH + H as a minor product \citep{vigren2010}. This means that, independently of the formation route of HCOOH, the initial cis/trans ratio of HCOOH remains uncertain.


In this work, we report a new gas-phase mechanism of destruction and post-formation of HCOOH that can explain the presence of c-HCOOH in the gas phase of cold dark clouds. According to this mechanism, either c-HCOOH and/or t-HCOOH react with HCO$^{+}$, a very abundant molecule toward these cold sources, yielding the three isomers of the cation HC(OH)$_{2}^{+}$. The four isomers of HC(OH)$_{2}^{+}$ differ in the positions of the hydrogens held by the oxygen atoms: one of the isomers shows the hydrogen atoms in cis position (cat4); a second one has the hydrogen atoms in trans position (cat3); and the third and fourth isomers have one hydrogen in cis and the other one in trans configuration (cat2 and cat1; see Figure 1).

Our results show that cat1 is the most abundant cation. In a second step, cat 1 reacts with NH$_{3}$ through different reaction channels leading back to c-HCOOH and t-HCOOH with a cis/trans ratio of 25.7\%. This novel mechanism would explain the presence of c-HCOOH in the gas phase of cold sources, and it would justify the higher abundances of t-HCOOH measured in these sources with respect to c-HCOOH. We note, however, that the discrepancy between the cis/trans ratio found in this work and the observations suggests that additional destruction mechanisms of the cations HC(OH)$_{2}^{+}$ might play an important role. The proposed sequential acid-base (SAB) mechanism explains the cis-trans interconversion in cold dark clouds, even when the direct reaction is endothermic. Since this mechanism can also operate for other molecules, we propose that it should be considered for any study focusing on the isomerism of organic molecules in the ISM.

\section{Computational details}
\subsection{Electronic structure calculations}

All the geometries were optimized without constraints at the double-hybrid rev-DSD-PBEP86 functional \citep{kozuch11,santra19} with the dispersion-corrected D3(BJ) \citep{grimme10,grimme11} method in combination with the augmented Dunning's triple-zeta correlation-consistent basis set aug-cc-pVTZ \citep{dunning89,kendall92}. All the calculations carried out in this work involve cation-neutral interactions, thus, the choice of the method (rev-DSD-PBEP86-D3(BJ)) is based on its good ability to describe non-covalent interactions of the cation-neutral pair \citep{Spicher2021} as well as for kinetics and thermodynamics \citep{Goerigk2017}. All the stationary points were characterized by frequency calculations at the same level of theory, showing none and one imaginary frequencies for energy minima and transition structures respectively. We checked that each transition structure belonged to the correct reaction path by the corresponding intrinsic reaction coordinate analysis (IRC). Electronic energies ($E$) were refined by computing single points with the CCSD(T)-F12 method \citep{adler2007,knizia2009} on the rev-DSD-PBEP86-D3(BJ)/aug-cc-pVTZ geometries in combination with the auxiliary and special orbital basis set cc-pVTZ-F12-CABS and cc-pVTZ-F12, respectively. Finally, anharmonic ZPE ($ZPE_{Anh}$) corrections were evaluated at the rev-DSD-PBEP86-D3(BJ)/aug-cc-pVTZ level within vibrational perturbative theory to second order VPT2 \citep{Barone2004}.

All the calculations with the double-hybrid and with the coupled cluster method were carried out with the Gaussian 16 \citep{gaussian} and ORCA \citep{neese12,neese20} software packages respectively. The images of the structures were done with the Cylview software \citep{cylview}.

\subsection{Downhill MEP calculations}

For all barrierless processes we computed the minimum energy path (MEP) downhill calculations. Starting from the equilibrium geometries of both fragments at a distance between 10 and 12 angstrom, we searched all the possible MEPs by rotating the equilibrium geometries of both fragments and then searching the MEP for all the input geometries. The initial geometries were generated by randomly sampling the Euler angles of the HCO$^{+}$ fragment with respect to the acid molecule frame, with a distance between the center of masses randomly located in the 10-12 angstrom interval.

For each association we ran 100 MEPs in order to cover all the possible orientations between the two fragments. For these calculations we used the rev-DSD-PBEP86-D3(BJ) method in combination with the cc-pVDZ basis set. The results show that all the potentials converge to the same path below 8 angstrom. That is, the reorientation of the fragments until they acquire a position similar to that of the minimum energy geometry is achieved in about 3-4 angstroms. The increase in the basis functions neither change the shape of the potentials nor the ratio found between competitive MEPs. 

\subsection{Kinetics calculations}
\subsubsection{Barrierless association reactions}

The rate constants for all barrierless bimolecular association reactions were computed within the phase space theory (PST) \citep{pechukas1965detailed,chesnavich1986multiple}. The attractive potentials between both fragments is described by a \(V_{MEP} = \frac{-C{_n}}{R^n}\) functional form. The $C{_n}$ constant is the long-range coefficient and is obtained from a fit of the rev-DSD-PBEP86-D3(BJ)/cc-pVDZ energies. The energies used to carry out the fitting were those obtained from any of the 100 MEPs for each association starting at distances of about 8.5 angstroms. The variable $R$ corresponds to the nuclear positions of the two atoms that are closer to each other along the MEP. We checked which functional form fits better to the potentials computed at rev-DSD-PBEP86-D3(BJ)/cc-pVDZ varying $n$ from 2 to 6. The best fittings were obtained for $n$ = 3. Thus, the functional form that better describe the interaction is \(V_{MEP} = \frac{-C{_3}}{R^3}\). Considering that the rate-determining step for the viable reactions found here are the initial barrieless association reactions, we compared our global rate constants with those obtained from the Su-Chesnavich formula in order to validate our PST calculations. The comparison between the two kinetic treatments yields consistent results as reported in Appendix A.

\subsubsection{Global rate constants}

For elementary steps associated with well-characterized transition structures, unimolecular rate constants were computed using Rice-Ramsperger-Kassel-Marcus (RRKM) theory within the rigid-rotor harmonic-oscillator (RRHO) approximation \citep{ChemKin}. For each step, one-dimensional tunnelling corrections were computed using the Eckart model \citep{eckart1930penetration}. The comparison of the rate constants without considering quantum tunnelling leads to identical results since this effect is negligible for submerged barriers in the low pressure limit. The imaginary frequencies of the transition structures showing energy barriers above the reactants are indeed low, which generates wide potentials that make the tunnelling ineffective. 

Pressure and temperature-dependent rate constants were calculated by resolving the one-dimensional master equation using the MESS software \citep{georgievskii2013reformulation}. The global rate constants were computed in the 15-400 K temperature range and at a pressure of 1x10$^{-7}$ atm. Although herein we are interested in the low temperate range typical of the molecular clouds B5 and L483, we also computed rate constants up to 400 K with the aim of giving general and valuable kinetic information for the wide range of conditions of the ISM.

To describe the temperature dependence of the global rate constants as well as for giving valuable kinetic data for astrochemical codes we used the  Arrhenius-Kooij formula \citep{kooij}:

\begin{equation}
k(T) = \alpha \left(\frac{T}{300}\right)^\beta exp\left(-\frac{\gamma}{T}\right)
\label{AK}
\end{equation}

where \(\alpha\), \(\beta\) and \(\gamma\) are fitting parameters, determined using the computed rate coefficients at different temperatures.

\section{Results}
\label{results} 
\subsection{Reaction of c-HCOOH and t-HCOOH with HCO$^{+}$}

\subsubsection{Reaction mechanism}

In this work we have explored the reactivity of c-HCOOH and t-HCOOH with \textbf{HCO$^{+}$} with quantum-chemical calculations. The reaction mechanism shows that there are five exit channels distributed in the formation of cat1, cat2, and cat3, cat4, and the complex CO-H$_{3}$O$^{+}$ (+ CO) (see Figure 1).

\begin{figure}[h!]
  \centering
  \includegraphics[width=8cm,height=5.5cm]{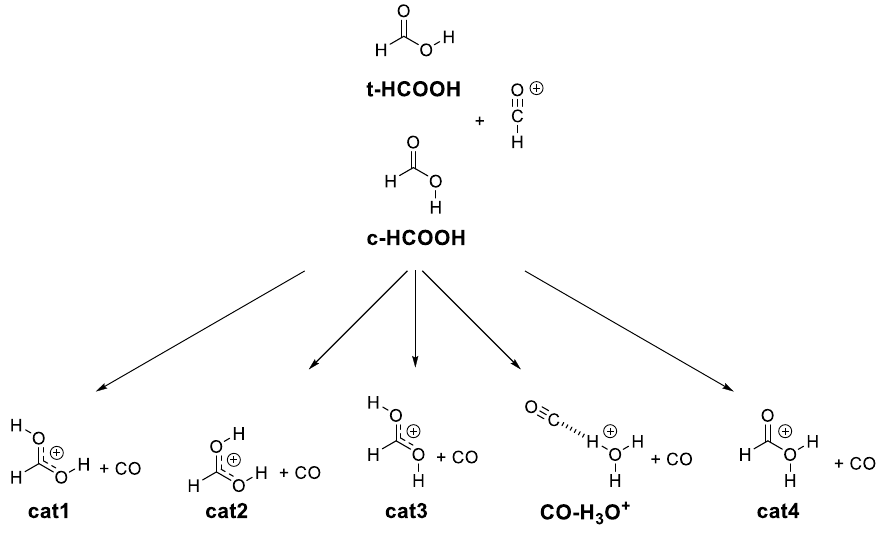}
  \caption{Protonation of t-HCOOH and c-HCOOH with HCO$^{+}$ to give the cations cat1, cat2, cat3, cat4 and CO-H$_{3}$O$^{+}$ (+ CO).}
\end{figure}

Considering that HCO$^{+}$ is a strong Bronsted acid with an electrophilic point in the carbon atom, and that HCOOH has three lone pairs located in the oxygen atoms (two in the oxygen of the carbonyl group and one aditional in the oxygen-holding hydrogen), six possible association reactions are possible for each isomer of HCOOH. 77 out of the 100 MEPs computed for the barrierless assotiation between HCO$^{+}$ and c-HCOOH lead to a hydrogen donnation to the oxygen of the carbonyl group forming the cat3 complex with a hydrogen bonding to CO (c-Int2, see Figure 2). In 22 out of the 100 MEPs, the same occurs but forming a cat1 complex (c-Int3). The last MEP found is a nucleophilic attack of the carbonyl group to the carbon of HCO$^{+}$ (c-Int1).

For t-HCOOH, we found four barrierless association reactions. 93 out of 100 MEPs lead to t-Int2, 3 to t-Int1, 2 to t-Int3 and 2 to t-Int4. From the intermediates t-Int2, t-Int3, c-In2 and c-Int3, we obtain four out of the five final exit channels through barrierless dissociation reactions. The intermediates formed after the proton donation of HCO$^{+}$ are the most stable with respect to c-HCOOH, being -50.82, -52.25, -48.84 and -50.18 kcal mol$^{-1}$ for t-Int2, t-Int3, c-Int2 and c-Int3, respectively. The intermediates formed after the electrophilic attack of HCO$^{+}$ are less stable than the previous ones, having energies -15.15, -37.51 and -36.08 kcal mol$^{-1}$ for t-Int4, t-Int1 and c-Int1, respectively.

The reaction mechanism studied for this reaction shows five additional intermediates formed from the pre-reactive complexes (t-Int6, 2CO-H$_{3}$O$^{+}$, t-Int5, c-Int4 and c-Int5). All of them are interconnected with each other through isomerization reactions (there are fifteen transition structures, see Figure 2). t-Int5 is linked to c-Int1, c-Int3, t-Int5 and t-Int2 through the transition structures ct-TS1, ct-TS3, t-TS3 and t-TS1. ct-TS3 shows a high energy barrier which is 9.09 kcal mol$^{-1}$ above the reactants, being a non-viable isomerization in the conditions of the ISM. ct-TS1, t-TS3 and t-TS1 show submerged barriers (-21.94, -26.45 and -27.81 kcal mol$^{-1}$ respectively). The only transition structure connecting t-Int4 with other intermediate is t-TS4 (-14.24 kcal mol$^{-1}$), for which there is a proton transfer yielding t-Int3. t-Int2 is also connected to t-Int3 through the transition structure c-TS3, which is just 3.71 kcal mol$^{-1}$ below the reactants. t-Int2 also leads to t-Int5 through t-TS2 and to c-Int2 through ct-TS2 (-35.71 kcal mol$^{-1}$), which corresponds to a torsion of the O-C-O-H. Both, c-Int2 and t-Int2 lead to the products cat3 and cat1, which are 38.43 and 40.08 kcal mol$^{-1}$ below the reactants.

The intermediate t-Int3 yields the complex 2CO-H$_{3}$O$^{+}$ through t-TS5. The latter intermediate loses one molecule of CO to give the complex CO-H$_{3}$O$^{+}$ + CO. These products are 37.70 kcal mol$^{-1}$ below the reactants. It should be noted that the majority of the energy in this process is likely dissipated by the increase in the translational energy of the products. In this way, the complex CO-H$_{3}$O$^{+}$ is likely a long-lived intermediate in the ISM since the complete dissociation into H$_{3}$O$^{+}$ and CO is 14.18 kcal mol$^{-1}$ above the bimolecular exit channel (for the unimolecular transformations, see Figure 3).

The rotation of the O-C-O-H bond of t-Int3 leads to c-Int5 through ct-TS6 (-28.45 kcal mol$^{-1}$). Both intermediates, t-Int3 and c-Int5 yield the products cat4 and CO (-21.13 kcal mol$^{-1}$). cat4 is a strong complex between H$_{2}$O and HCO$^{+}$ that could continue evolving into H$_{2}$O and HCO$^{+}$ and/or H$_{3}$O$^{+}$ and CO (Figure 3). The evolution of cat4 has been previously explored from the theoretical point of view \citep{Sekiguchi2004}. Our new and more accurate results (Figure 3) show that the unimolecular isomerization of cat4 into cat1 shows a transition structure with a very high energy barrier (9.00 kcal mol$^{-1}$ above the reactants) which cannot be overcome in the conditions of the ISM. cat 4 could also yield the complex CO-H$_{3}$O$^{+}$ through an energy barrier which is 20.24 kcal mol$^{-1}$ below the reactants. This process could occur in the conditions considered here if the products (cat4 + CO) are not completely stabilised (the barrier is just 0.89 kcal mol$^{-1}$ from cat4, see Figure 3). To verify that this last process is possible, high-level quantum dynamics calculations are required, which is out of the scope of this work.

Finally, moving to the intermediates formed from c-HCOOH, c-Int1 leads both, to c-Int3 through c-TS1 (-27.20 kcal mol$^{-1}$) and to c-Int4 through c-TS4 (-26.29 kcal mol$^{-1}$). As for t-Int2, the dissociation of c-Int3 leads to cat1. c-Int3 can also isomerize to c-Int4 through c-TS2. The latter could go back to c-Int1 or yield the intermediate t-Int5.

We note that all stationary points are connected to each other and thus, the isomerization from c-HCOOH to t-HCOOH can proceed with the latter behaving as the exit channel. The inverse isomerization is not possible under the conditions of the ISM since it is an endothermic process (4.02 kcal mol$^{-1}$).

\begin{figure*}[h!]
  \centering
  \includegraphics[width=18.5cm,height=15cm]{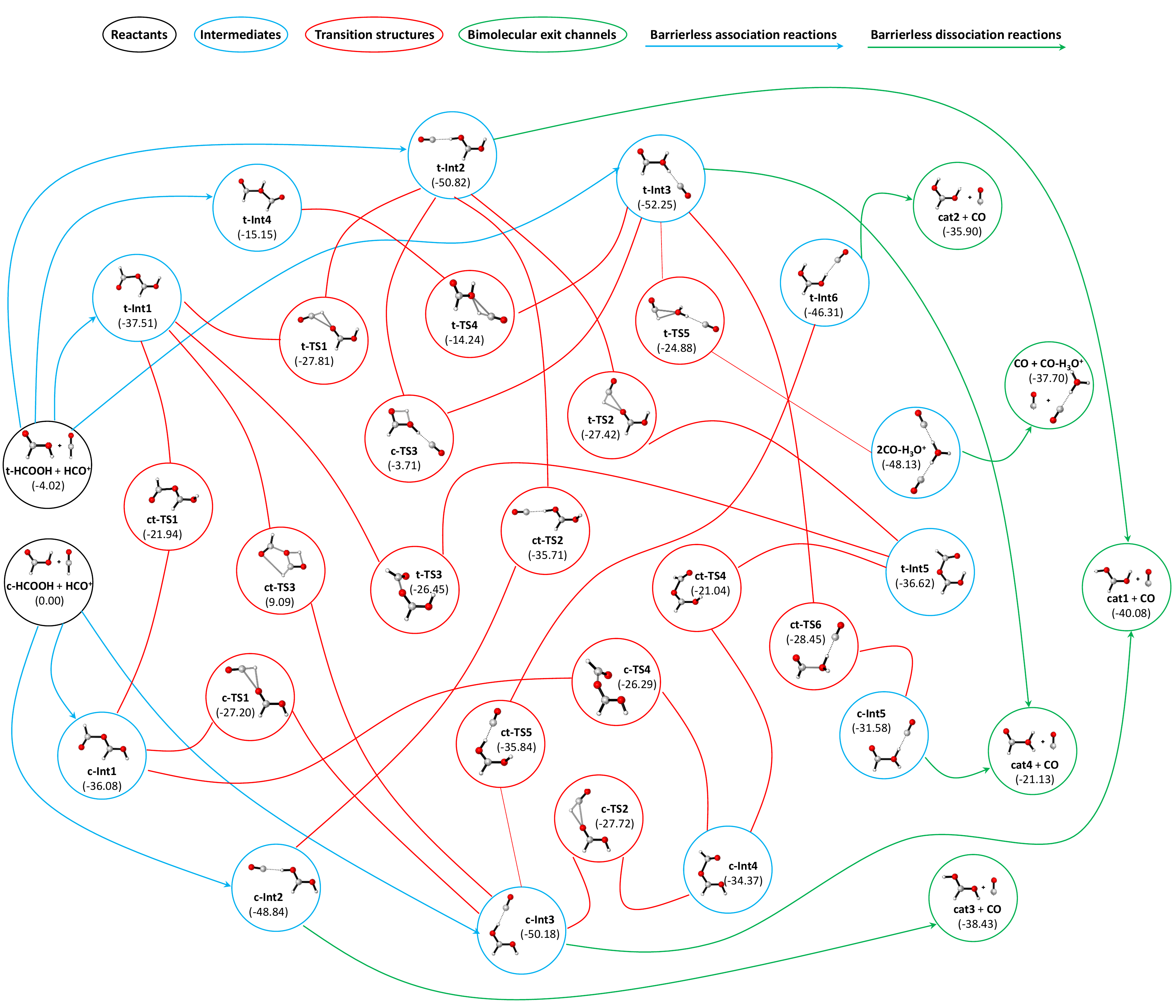}
  \caption{$E$+$ZPE$$_{Anh}$ energy profiles (in kcal mol$^{-1}$) for the reactions of c-HCOOH and t-HCOOH with HCO$^{+}$. The meaning of the circles and arrows are indicated in the legend at the top of the figure.}
\end{figure*}

\begin{figure}[h!]
  \centering
  \includegraphics[width=8.5cm,height=8.5cm]{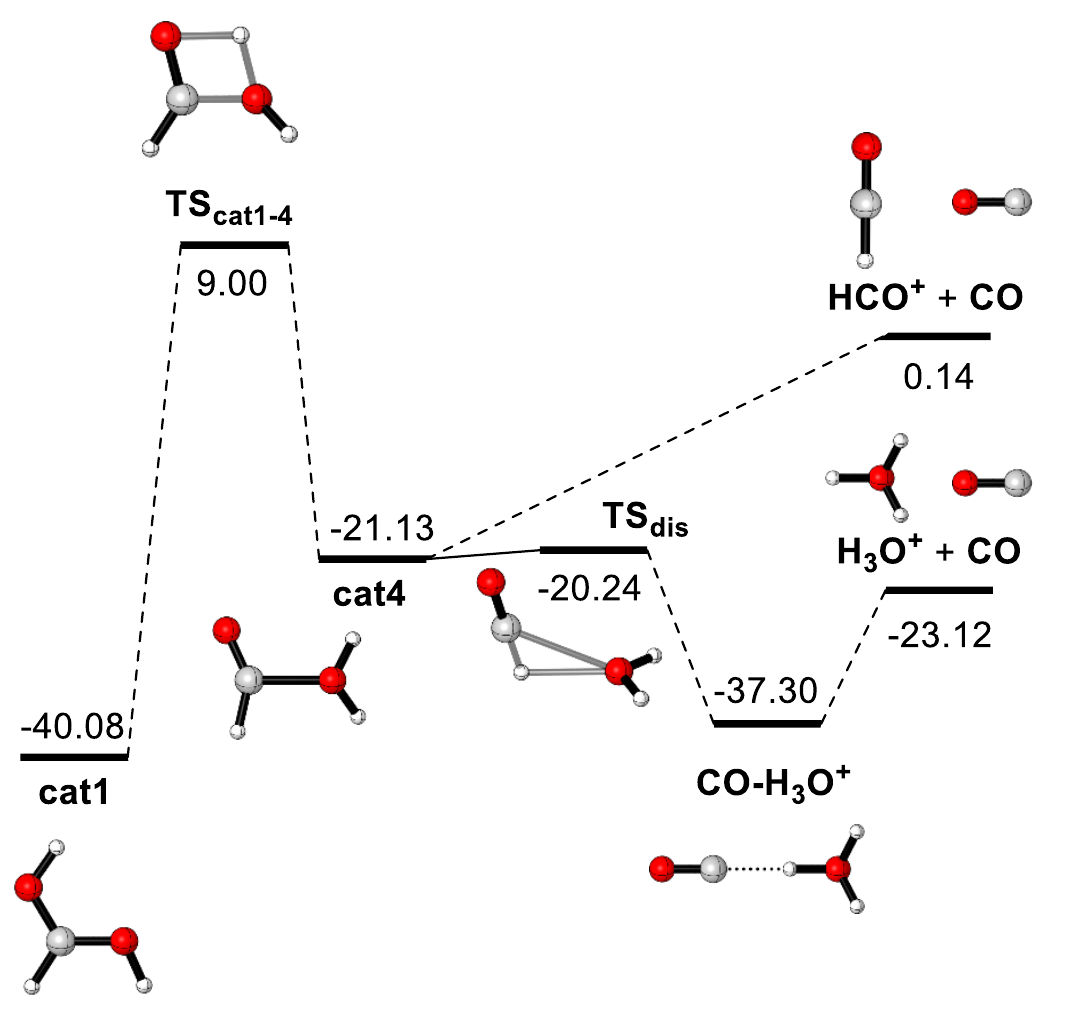}
  \caption{$E$+$ZPE$$_{Anh}$ energy profiles (in kcal mol$^{-1}$) for the unimolecular decomposition of cat1. Relative energies are given with respect to c-HCOOH and HCO$^{+}$.}
\end{figure}

\subsubsection{Kinetics}

Figure 4 shows the global rate constants for the destruction of c-HCOOH (blue curves) and t-HCOOH (orange curves). Figure 4 a) shows the destruction of c-HCOOH and t-HCOOH to give cat1 and CO. Both reactions show a non-Arrhenius behaviour in the whole range of temperatures which is the typical trend for a barrierless association between a cation and a neutral species. Both reactions are very fast processes, of the order of 5.2-6.2x10$^{-9}$ cm$^{3}$molecule$^{-1}$s$^{-1}$ at 15 K. The destruction of c-HCOOH is slightly faster than that of the t-HCOOH in the whole range of temperatures. Considering all possible pathways investigated here, the formation of cat1 from c-HCOOH and t-HCOOH are the fastest routes of the reaction mechanism HCOOH + HCO$^+$ (see below).  

In contrast to the formation of cat1, the formation of cat2 from c-HCOOH (Figure 4 b)) shows a non-Arrhenius behaviour only up to $\sim$260 K. This is because the rate of the reactions at very low temperatures is limited by the capture rate constants. Above $\sim$260 K the energy barriers increase their height and the temperature-dependent rate constants adopt an Arrhenius behaviour. This kinetic behaviour is also observed for the formation of cat2 from t-HCOOH (Figure 4 f)), but the transition between the non-Arrhenius to Arrhenius behaviour is reached at $\sim$170 K. In this case, also the formation of the product (cat2) is faster from c-HCOOH than from t-HCOOH, showing a global rate constant of 4.57x10$^{-11}$ and 1.13x10$^{-13}$  cm$^{3}$molecule$^{-1}$s$^{-1}$ at 15 K respectively.

The same temperature-dependence is observed for the formation of cat3 from t-HCOOH (Figure 4 g)), which is a faster process than the formation of cat2 from the same reactants, i.e. t-HCOOH + HCO$^{+}$. On the other hand, the formation of the same product from c-HCOOH shows a non-Arrhenius behaviuor (Figure 4 c)) for the whole range of temperatures since the rate is limited by the centrifugal barrier at the entrance channel, being the third fastest process of the whole mechanism, and having a global rate constant of 2.72x10$^{-9}$  cm$^{3}$molecule$^{-1}$s$^{-1}$ at 15 K.

The plot d) of Figure 4 reports the rate coefficients for the formation of cat4 + CO (solid blue line) and CO-H$_{3}$O$^{+}$ + CO (dotted blue line). These reactions are the slowest of the whole process since the system has to reach t-Int3 from c-HCOOH + HCO$^{+}$ . In these pathways the system should pass through the transition state c-TS3, which shows the highest submerged barrier of the whole mechanism (-3.73 kcal mol$^{-1}$). The rate of these reactions decreases fast up to $\sim$60 K since a slightly increase in the temperature makes the energy barrier to increase, approaching the energy of the reactants. After this inflection point, the rate constants show an Arrhenius behaviour because the system starts to have thermal energy for overcoming the energy barrier. On the other hand, the rate of the formation of these products from t-HCOOH + HCO$^{+}$ are fast processes (see Figure 4 h)) since, although the system should also pass throughout t-Int3, the formation of the latter intermediate comes from a barrierless process. Then, the global rate coefficients are limited by the capture rate constants. The formation of cat4, albeit less stable than CO-H$_{3}$O$^{+}$ by 16.57 kcal mol$^{-1}$, is faster because there are two exit channels, one of which is a barrierless process. In contrast, for the formation of CO-H$_{3}$O$^{+}$ it is mandatory to pass through a transition state (t-TS5).

As commented in the previous section, all intermediates are directly or indirectly connected in this reaction, so the formation of t-HCOOH + HCO$^{+}$ starting from c-HCOOH + HCO$^{+}$ could occur in the conditions of the ISM. However, it is one of the slowest processes for this reaction. The reverse process is not viable in the ISM since it is endothermic.

In summary, the formation of cat1 is the most favoured pathway starting from both c-HCOOH and t-HCOOH, showing the highest product branching ratio for the whole range of temperatures (15-400 K) (see Appendix B). On the contrary, the formation of cat2 and cat3 covers the remaining of the product branching ratio when the reactant is c-HCOOH. Indeed, the formation of cat4 and the complex CO-H$_{3}$O$^{+}$ is negligible. When the reactant is t-HCOOH, however, there is not any formation pathway of cat2 but the product branching fractions for cat4 and the complex CO-H$_{3}$O$^{+}$ are noticeable. From all this, it is likely that HCOOH is  converted completely to its cations in molecular dark clouds as a consequence of the high rate for the formation of cat1, cat2 and cat3 and the high abundance of HCO$^{+}$ in B5 and L483.

\subsection{Reaction of HC(OH)$_{2}^{+}$ (cat1) with NH${_3}$}

\subsubsection{Reaction mechanism}

Ammonia (NH$_{3}$) is a very abundant species in B5 and L483, with a relative abundance with respect to H$_2$ as high as the electron abundance \citep{agundez2013}. Considering that most HCOOH should be converted to cat1 in molecular dark clouds (see Section 3.1), in this Section we study the destruction of the latter cation with NH$_{3}$. 

The structure of cat1 presents two acidic hydrogens nearly equivalent. The capture of the proton in cis position of cat1 by NH$_{3}$ would lead back to t-HCOOH whereas the capture of the proton in trans position would lead to c-HCOOH. 

Continuing with the methodology mentioned in section 2.2, we calculate the MEPs for 100 different orientations of cat1 and NH${_3}$ in order to explore all possible entrance channels. We found three barrierless association reactions that are summarised in Figure 5. 65 out of 100 MEPs lead to Int-H3, which is a hydrogen-bonding complex between t-HCOOH and amonium ion (NH$_{4}^{+}$). 26\% of the MEPs yield the same complex but with c-HCOOH (Int-H2). Finally, 9\% of MEPs lead to the intermediate Int-H1. For the latter, there is a non-covalent interaction between the lone pair of ammonia and the hydrogen held by the carbon atom of cat1. The formation of this complex does not involve any breaking or forming bond. It is important to highlight that as soon as Int-H1 is formed, it isomerises to Int-H3 by overcoming a very low barrier (see below). Therefore, 74\% of the trajectories end up in Int-H3.

\begin{figure*}[h!]
  \centering
  \includegraphics[width=18cm,height=12.5cm]{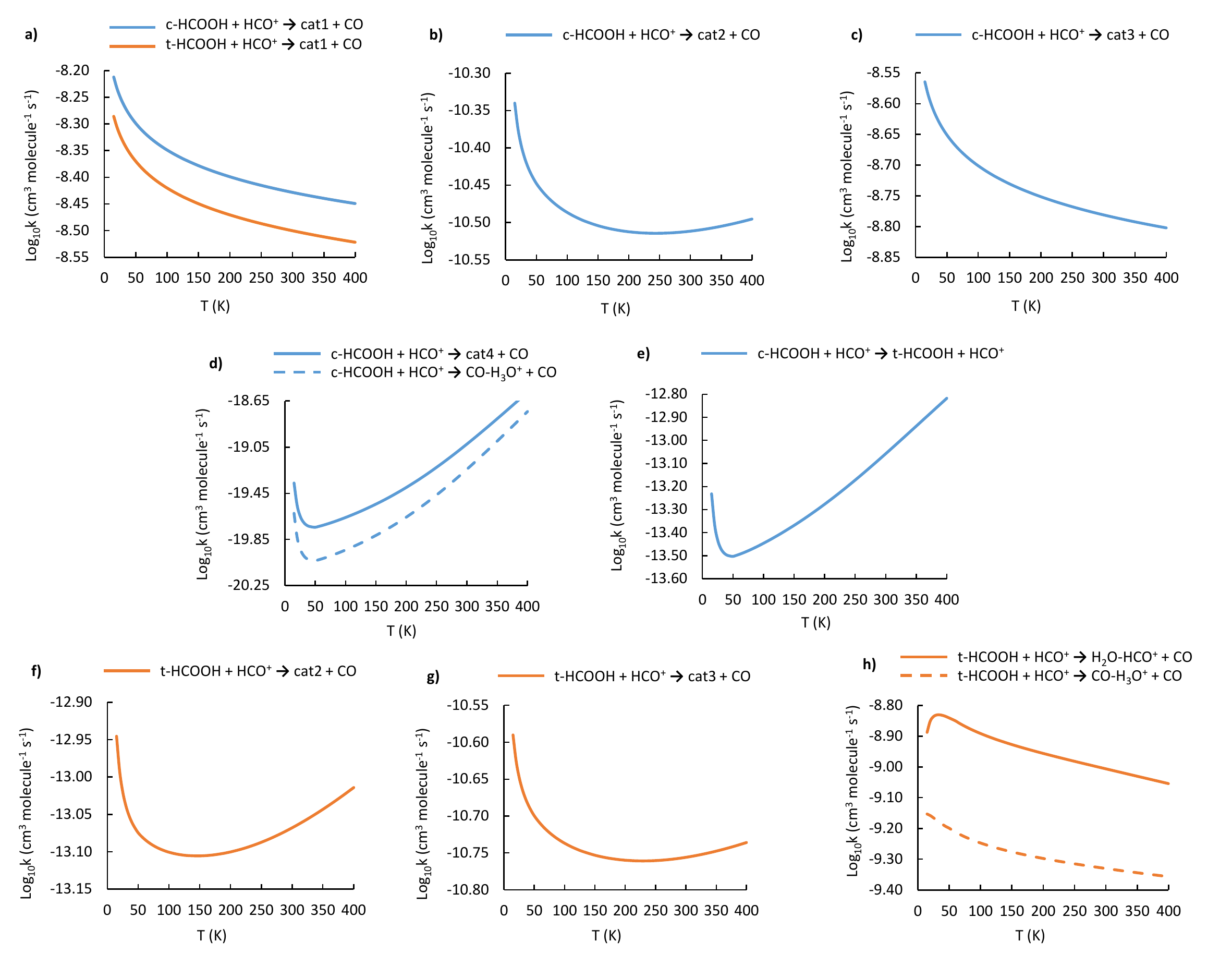}
  \caption{Plots of the global rate coefficients (logarithmic scale) for the destruction of c-HCOOH and t-HCOOH with HCO$^{+}$ in the 15-400 K range.}
\end{figure*}

\begin{figure}[h!]
  \centering
  \includegraphics[width=7.5cm,height=8cm]{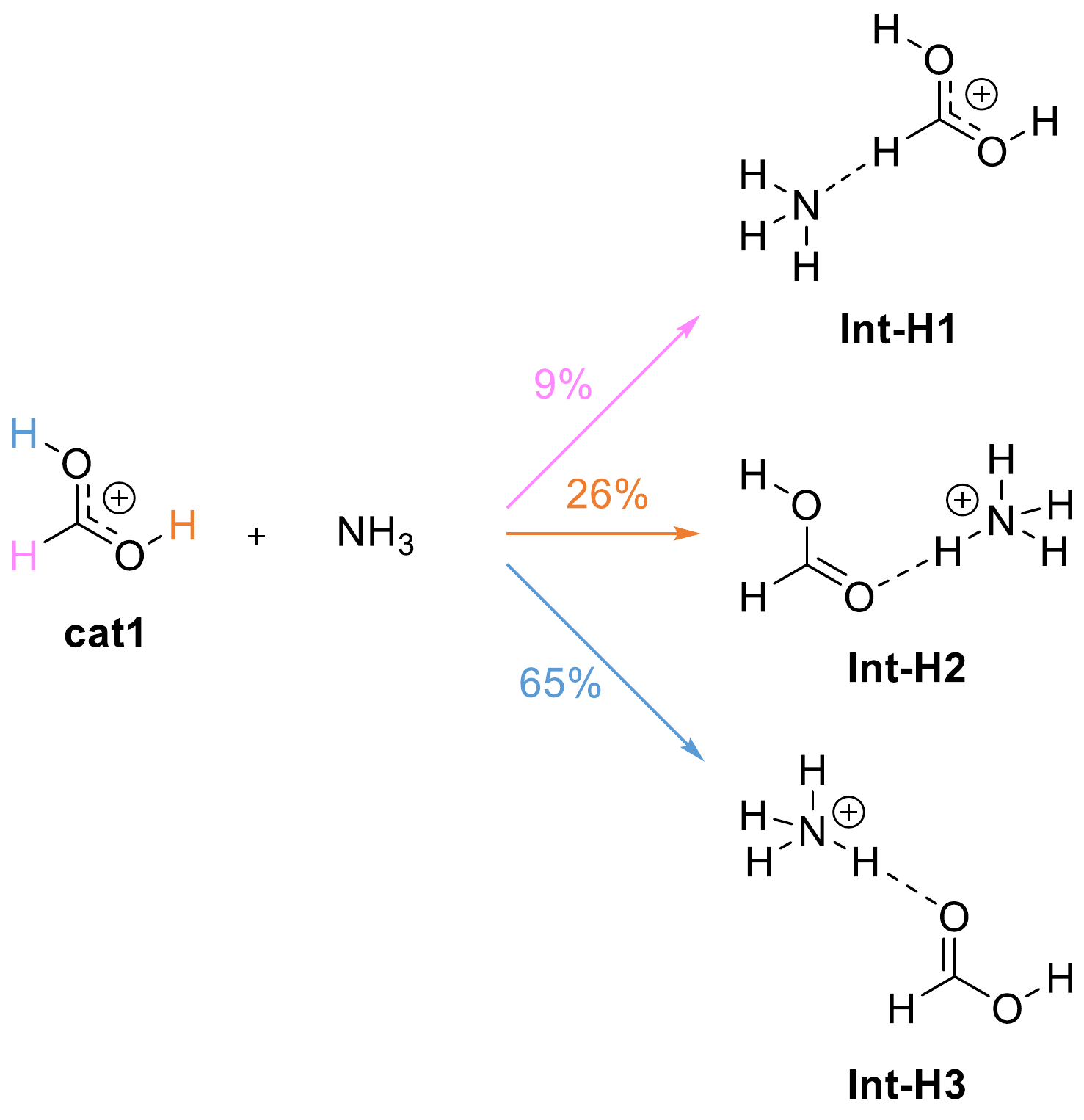}
  \caption{Branching ratios of the 100 MEP computed for the barrierless association of cat1 with NH$_{3}$.}
\end{figure}

\begin{figure*}[h!]
  \centering
  \includegraphics[width=18cm,height=9cm]{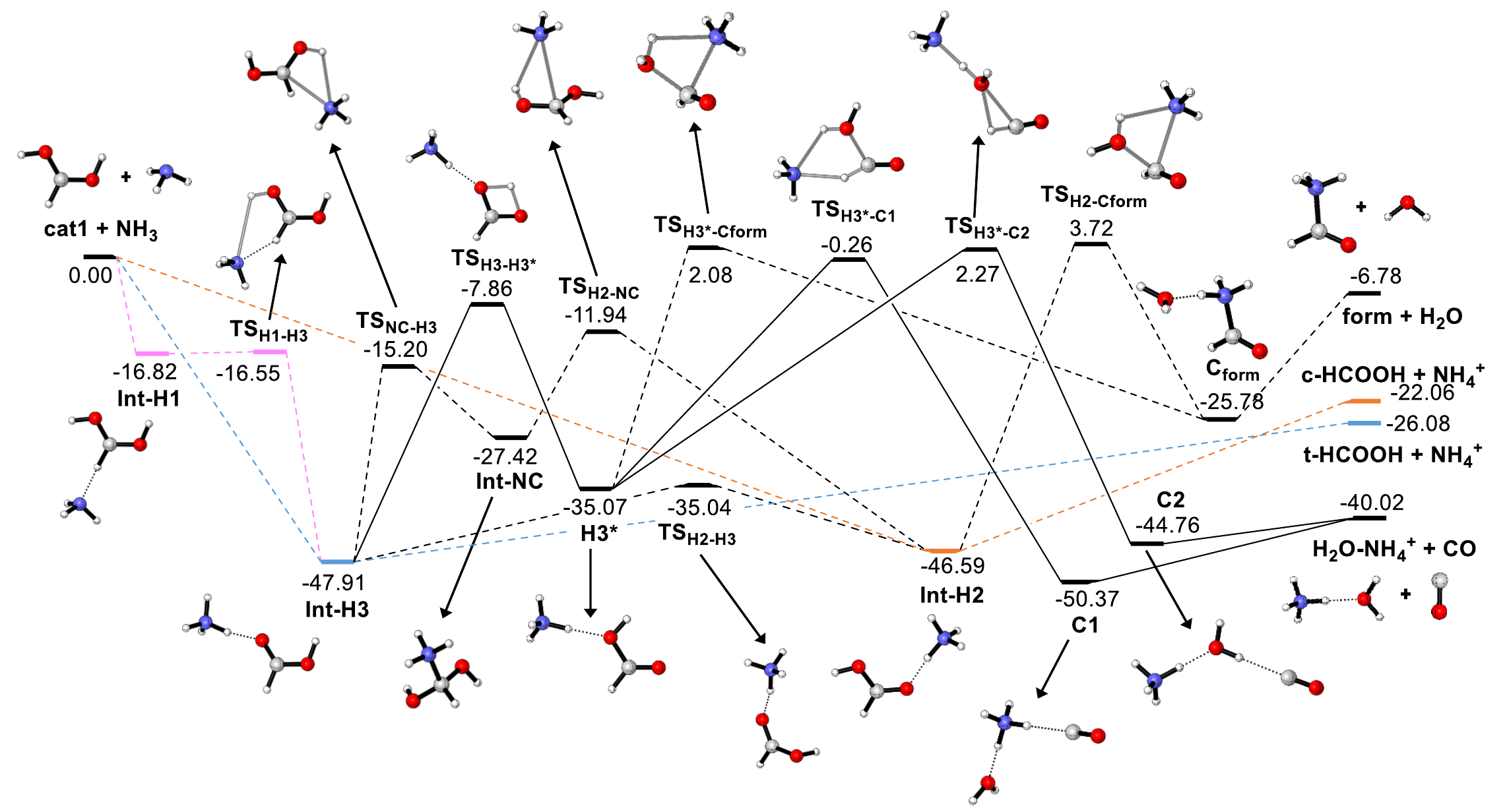}
  \caption{$E$+$ZPE$$_{Anh}$ energy profiles (in kcal mol$^{-1}$) for the reaction of cat1 (HC(OH)$_{2}^{+}$) with NH$_{3}$.}
\end{figure*}

\begin{figure*}[h!]
  \centering
  \includegraphics[width=18cm,height=6.2cm]{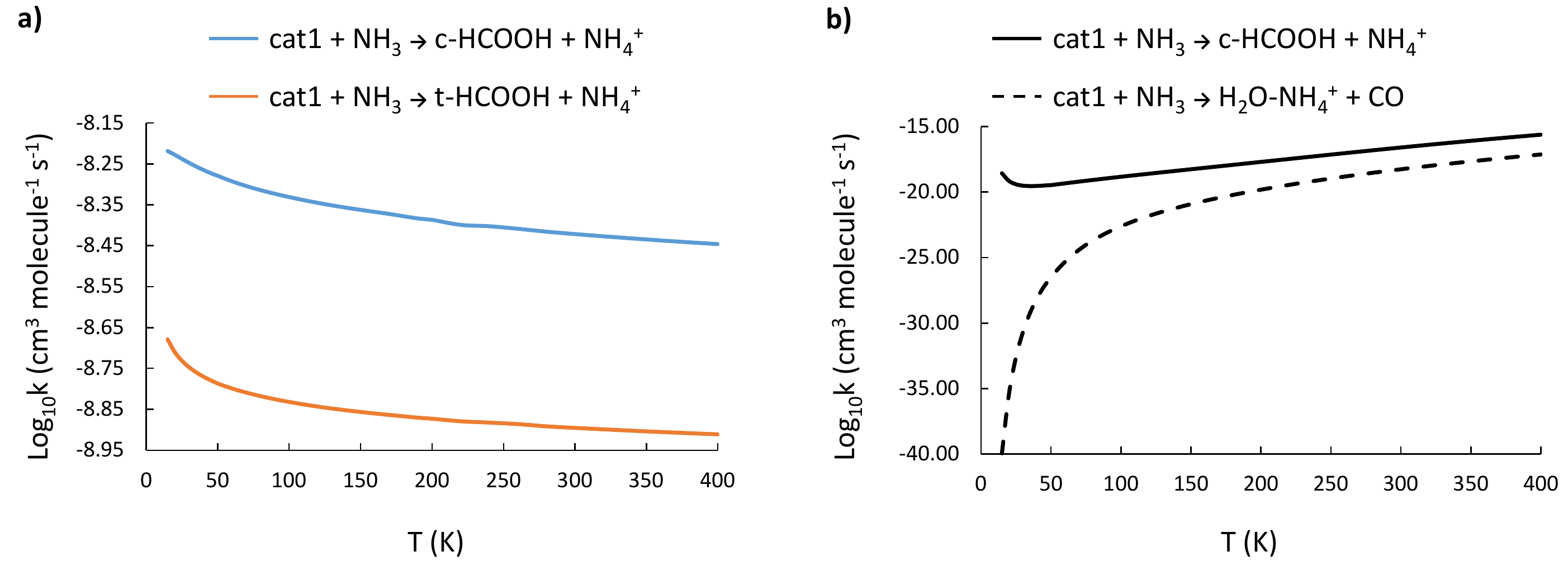}
  \caption{Global rate coefficients (in logarithmic scale) for the destruction of cat1 (HC(OH)$_{2}^{+}$) with NH$_{3}$ in the 15-400 K range.}
\end{figure*}

Figure 6 shows the full $E$+$ZPE_{Anh}$ energy profiles for the destruction of cat1 by reaction with NH${_3}$ leading back to c-HCOOH and t-HCOOH and ammonium ion (NH$_{4}^{+}$); N-protonated formamide (form) and H$_{2}$O; and the complex H$_{2}$O-NH$_{4}^{+}$ and CO. The blue profile of Figure 6 shows the direct formation and post dissociation of the complex Int-H3 leading to t-HCOOH and NH$_{4}^{+}$. The orange profile shows the same but for the formation of the intermediate Int-H2 and the post dissociation to c-HCOOH and NH$_{4}^{+}$. The pink profile corresponds to the formation of the Van der Waals complex Int-H1, which is 29.77 kcal mol$^{-1}$ and 31.09 kcal mol$^{-1}$ above Int-H2 and Int-H3, respectively. The latter intermediates present an energy difference of just 1.32 kcal mol$^{-1}$, being the trans intermediate (Int-H3) the most stable. The unstable complex Int-H1 could dissociate back to the reactants or isomerize rapidly to Int-H3 through the transition structure TS$_\mathrm{H1-H3}$, which is just 0.27 kcal mol$^{-1}$ above Int-H1. 

As in the mechanism depicted in Figure 2, the stationary points of the mechanism shown in Figure 6 are connected through isomerization reactions between the intermediates Int-H3 and Int-H2. We found two isomerization mechanisms; TS$_\mathrm{H2-H3}$ correspond with the torsion of the H-O-C-O diedral angle of the formic acid moiety of the complexes Int-H3 and Int-H2, which have energy barriers of 12.87 kcal mol$^{-1}$ from Int-H3 and 11.55 kcal mol$^{-1}$ from Int-H2. The other isomerization is a stepwise mechanism where the NH$_{4}^{+}$ moiety of Int-H3 gives a proton to the oxygen of the carbonyl group whilst a new N-C bond is formed in a concerted manner. This transition structure (TS$_\mathrm{NC-H3}$), which shows an energy barrier of 37.71 from Int-H3, leads to the intermediate Int-NC. The evolution of Int-NC into Int-H2 through TS$_\mathrm{H2-NC}$ is similar to the previous (TS$_\mathrm{NC-H3}$) but involving the proton which is in the trans position, being 3.26 kcal mol$^{-1}$ above the latter. 

Once Int-H2 is formed, either from the reactants through a barrierless association reaction or from Int-H3 through isomerization reactions, it could evolve through the exit channel that leadS to c-HCOOH and NH$_{4}^{+}$ or isomerize into C$_\mathrm{form}$ through the transition structure TS$_\mathrm{H2-Cform}$. The latter is 3.72 kcal mol$^{-1}$ above the reactants, being a non-viable process under ISM conditions. In this transition structure there is a transfer of a proton from the NH$_{4}^{+}$ to the oxygen of the OH while the substitution reaction of NH$_{3}$ by OH forms concertedly the N-protonated formamide interacting with a water molecule (C$_\mathrm{form}$). The breaking of the hydrogen bond of the latter complex forms N-protonated formamide (HCONH$_{3}^{+}$, called herein form) and H$_{2}$O. 

On the other hand, the trans intermediate Int-H3 suffers other type of rearrangements where there is a hydrogen transfer from the OH to the oxygen of the carbonyl group in the formic acid moiety though the transition structure TS$_\mathrm{H3-H3*}$, which shows a high energy barrier from Int-H3 but still submerged with respect to the reactants (40.05 kcal mol$^{-1}$). This reaction pathway leads to H3* in which the NH$_{4}^{+}$ forms a hydrogen bonding with the OH group of t-HCOOH. This intermediate could evolve through three different reaction pathways. The first transition structure leads again to C$_\mathrm{form}$ through the transition structure TS$_\mathrm{H3*-Cform}$, whose geometry is similar to that of TS$_\mathrm{H2-Cform}$. The difference between both transition structures is the orientation of the hydrogen of the OH group. The energy of TS$_\mathrm{H3*-Cform}$ is also 2.08 kcal mol$^{-1}$ above the reactants, and thus, it is unlikely that leads to N-protonated formamide (form) in the ISM. 

Int-H3 could also isomerize into the complexes C1 and C2. Both complexes are two conformations of the same molecules. In C1, the NH$_{4}^{+}$ acts as a two hydrogen bonding donor of CO and H$_{2}$O whereas in C2 H$_{2}$O acts as hydrogen donor with CO and hydrogen bond acceptor with NH$_{4}^{+}$. The most energetically favoured situation is C1 (5.61 kcal mol$^{-1}$ below C2). In both complexes the weakest hydrogen bond is that formed with CO, and thus, they dissociate into the complex H$_{2}$O-NH$_{4}^{+}$ and CO. The formation of C2 from H3* shows a transition structure (TS$_\mathrm{H3*-C2}$) which is 2.27 kcal mol$^{-1}$ above the reactants whereas the formation of C1 from H3* pass through TS$_\mathrm{H3*-C1}$, which is just 0.26 kcal mol$^{-1}$ below the reactants. This situation makes this process viable at 0 K. However a small increase in the temperature consequently increases the  height of the energy barrier, becoming a non-viable process even at 15 K. This computational study shows that the only viable processes in the destruction of cat1 with NH$_{3}$ is the back formation of c-HCOOH and t-HCOOH (see below).

\subsubsection{Kinetics}

Figure 7 shows the plots of the rate constants for the destruction of cat1 described in Section 3.2.1. Figure 7 a) corresponds to the rate coefficient for the formation of c-HCOOH and NH$_{4}^{+}$ (blue line) and t-HCOOH and NH$_{4}^{+}$ (orange line). Both reactions present a non-Arrhenius behaviour since there is no energy barrier in the reaction pathways and the rate-determining step of the whole process is the first barrierless association reaction. This kinetic behaviour was also observed in the formation of cat1 (Figure 4 a)). The rate of formation of t-HCOOH is higher than that of c-HCOOH, varying approximately a factor 3 in the whole temperature range.

The dotted lines of the Figure 7 b) reports the rate coefficients for the formation of N-protonated formamide. This process shows a typical Arrhenius temperature dependence with very low rate constant values in the 15-400 K temperature range. This reaction is unlikely to occur in the ISM since its formation requires overcoming energy barriers that are above the energy of the reactants. Likewise, the formation of H$_{2}$O-NH$_{4}^{+}$ and CO is a non-viable process since the lowest energy barrier yielding the products is just 0.26 kcal mol$^{-1}$ below the reactants at 0 K. Thus, an increase in the temperature places the energy barrier above the reactants, becoming a non viable reaction under ISM conditions. This is the reason why the rate constant weakly increases at very low temperatures (see the black line in Figure 7 b)). 

This kinetic study demonstrates that the destruction of cat1 by collisions with NH$_{3}$ leads exclusively to c-HCOOH and t-HCOOH. The branching cis/trans ratio found between the rate coefficients for the formation of HCOOH is thus 25.7\% at 15 K.

\section{Discussion}

\subsection{Destruction of HCOOH with HCO$^{+}$}

\subsubsection{Product branching ratios}

Our new high-level $ab$ $initio$ results shows that the destruction of c-HCOOH, if present initially, and of t-HCOOH by reaction with HCO$^{+}$ is a very fast process (see Table 1 for the Arrhenius-Kooij parameters). These destruction mechanisms lead exclusively to the formation of the HCOOH cations cat1, cat2 and cat3 (see Table 3 for the product branching ratios). The other cation of HCOOH, cat4, has been previously considered in astrochemical studies \citep{vigren2010}. However, our kinetic study shows that cat4 should be formed through other chemical routes different from the one studied here since its destruction is likely to occur after its formation.

The branching ratios obtained with our kinetic study indicate that most c-HCOOH and t-HCOOH are transformed into the cation cat1. As mentioned in section 1, c-HCOOH may not be present initially since it is destroyed very fast on grains by reaction with hydrogen atoms \citep{Molpeceres2022} and in the gas phase by isomerisation reactions \citep{juan2021b}. Even so, c-HCOOH is detected in cold molecular dark clouds. Considering that HCO$^{+}$ is a very abundant molecule in these sources, c-HCOOH could be transformed in the gas phase after reacting with it. This reaction yields mainly to the formation of cat1 (\(\sim\)69\% of the total product branching ratio), cat3 (\(\sim\)30.6\%) and to a less extent, cat2 (\(\sim\)0.5\%) (see Table B.1). On the other hand, t-HCOOH, much more abundant than c-HCOOH, lead to cat1 and cat3 with a ratio of 71.9\% and 0.36\% at 15 K respectively. The rest is distributed in the formation of cat4 (9.76\%) and  CO-H$_{3}$O$^{+}$ (17.98\%) (see Table B.2). It should be noted that the formation of cat4 + CO is an exothermic process (-21.13 kcal mol$^{-1}$) in which most of the energy goes into increasing the velocity of the fragments. However, cat4 could evolve into CO-H$_{3}$O$^{+}$ through a transition structure that is just 0.89 kcal mol$^{-1}$ above the cation (cat4). Considering that the fragments could retain some internal energy, the life time of cat4 should be low and it is completely transformed into  CO-H$_{3}$O$^{+}$. This process has been demonstrated experimentally for the reaction of HCOOH with H$_{3}^{+}$ (see Section 4.1.2.).

\subsubsection{Comparison of the reactions HCOOH + HCO$^{+}$ and HCOOH + H$_{3}^{+}$}

It has been shown that the reaction of HCOOH with H$_{3}^{+}$ is a completely dissociative process that yields either H$_{3}$O$^{+}$ + CO or HCO$^{+}$ + H$_{2}$O under low-pressure conditions \citep{Mackay1978,Sekiguchi2004}. The first step of the reaction should be the formation of the cation cat1 and molecular hydrogen as well as for the reaction with HCO$^{+}$ is cat1-cat4 and carbon monoxide (CO). In the reaction with H$_{3}^{+}$ the total energy dissipation of the system through the increase of the translational energy of the fragments is not enough as to thermalize the products (cat1 + CO). This is due to the reaction HCOOH + H$_{3}^{+}$ being highly exothermic and thus the remaining internal energy in the fragments allows cat1 to isomerize into H$_{3}$O$^{+}$ + CO and HCO$^{+}$ + H$_{2}$O \citep{Mackay1978}. This process is however not observed for the reaction HCOOH + HCO$^{+}$ because the formation of cat1 + CO is not so exothermic as to keep the fragments as excited as to isomerize to cat4, being a reaction without detectable dissociation \citep{Mackay1978}. The energy of the transition structure that connect cat1 to cat4 is 9.00 kcal mol$^{-1}$ above the reactants (c-HCOOH + HCO$^{+}$), which makes it a non-viable process under the conditions considered here (see Figure 3). Therefore, this indicates that cat1 should form in dark molecular clouds.

\begin{table*}[h!]
\setlength{\tabcolsep}{1pt} 
\caption{Fitting parameters for the destruction of c-HCOOH and t-HCOOH by reaction with HCO$^{+}$ for the temperature range indicated in parenthesis.}  
\label{tab1}      
\centering           
\begin{tabular}{c c c c }  
\hline\hline    
\multicolumn{4}{c}{c-HCOOH + HCO$^{+}$}\\
\hline
 Products & $\alpha$ (cm${^3}$ molecule$^{-1}$ s$^{-1}$) & $\beta$ & $\gamma$ (K)\\
\hline         
t-HCOOH + CO (15-120 K) & 5.21x10$^{-14}$ & 0.60 & -28.90\\
t-HCOOH + CO (130-400 K) & 2.29x10$^{-14}$ & 3.01 & -405.93\\
cat1 + CO (15-400 K) & 3.73x10$^{-9}$ & -0.17 & 0.03\\
cat2 + CO (15-120 K) & 2.88x10$^{-11}$ & -8.43x10$^{-2}$  & -3.12\\
cat2 + CO (130-400 K) & 2.57x10$^{-11}$ & 0.25 & -55.19\\
cat3 + CO (15-400 K) & 1.66x10$^{-9}$ & -0.17 & 0.08\\
cat4 + CO (15-120 K) & 3.81x10$^{-20}$ & 0.88  & -41.44\\
cat4 + CO (130-400 K) & 4.45x10$^{-21}$ & 6.17 & -921.80\\
CO-H$_{3}$O$^{+}$ + CO (15-120 K) & 2.10x10$^{-20}$ & 0.95 & -44.72\\
CO-H$_{3}$O$^{+}$ + CO (130-400 K) & 1.70x10$^{-21}$ & 7.00 & -1.06x10$^{2}$\\
\hline        
\multicolumn{4}{c}{t-HCOOH + HCO$^{+}$}\\
\hline 
cat1 + CO (15-400 K) & 3.16x10$^{-9}$ & -0.17 & 0.23\\
cat2 + CO (15-120 K) & 7.50x10$^{-14}$ & 7.90x10$^{-3}$  & -6.60\\
cat2 + CO (130-400 K) & 5.92x10$^{-14}$ & 0.70 & -113.18\\
cat3 + CO (15-120 K) & 1.63x10$^{-11}$ & -7.30x10$^{-2}$  & -3.51\\
cat3 + CO (130-400 K) & 1.44x10$^{-11}$ & 0.29 & -60.80\\
cat4 + CO (15-400 K) & 1.02x10$^{-9}$ & -0.31 & 10.41\\
CO-H$_{3}$O$^{+}$ + CO (15-400 K) & 4.70x10$^{-10}$ & -0.19 & 2.70\\
\hline
\end{tabular}
\end{table*}

\subsection{Destruction of HC(OH)$_{2}^{+}$ (cat1) with NH$_{3}$: The origin of c-HCOOH in dark molecular clouds}

It has been proposed that the formation of HCOOH must occur in dust grains by hydrogenation of the HOCO radical \citep{Ioppolo2010,taquet17}. Recently, \citet{Molpeceres2022} demonstrated that whatever the cis/trans ratio of HCOOH is, c-HCOOH should not withstand the conditions on the surface of dust grains since its destruction by H atoms is a very fast process. In the gas phase, isomerisation transformations should yield predominantly t-HCOOH, which does not explain the presence of c-HCOOH of 5\% with respect to t-HCOOH in the gas phase of cold dark clouds \citep{juan2021b}.

In this work, we have shown that t-HCOOH and c-HCOOH can  react by collisions with HCO$^{+}$ forming efficiently cat1, cat2 and cat3, being cat1 the most abundant product. This highly reactive species can be destroyed back to c-HCOOH and t-HCOOH by reaction with NH$_{3}$ (see Section 3.2). To study the destruction of cat1, we selected NH$_{3}$ since it is one of the most abundant molecules in molecular dark clouds such as B5 or L483. Our kinetic results demonstrate that the formation of c-HCOOH and t-HCOOH are the only products. Indeed, N-protonated formamide (HCONH$_{3}^{+}$) + H$_{2}$O and the complex H$_{2}$O-NH$_{4}^{+}$ + CO are not formed through this chemical process.

Tables 2 and 3 report respectively the Arrhenius-Kooij parameters and the branching ratios for the destruction of cat1, being practically constant throughout the whole temperature range. At $T_{\rm kin}$ of \(\sim\)15 K the ratio of c-HCOOH with respect t-HCOOH is 25.7\%. It should be noted that this process is cyclic, that is, HCOOH is destroyed by HCO$^{+}$ and then HCOOH is again formed by the destruction of cat1, giving rise to a constant cis/trans ratio over time. 

Although not in the observed amount with respect to t-HCOOH, our study provides an origin for the presence of c-HCOOH under the cold temperatures of molecular dark clouds. The expected ratio is 25.7\%, far from the one reported by observations (6\%). This difference could be due the destruction of both cat2 and cat3, which should also be considered. In addition, note that there could be other mechanisms for the destruction of the cations with abundant species acting as base such as CO, OH and H$_{2}$O among others. 

The most intuitive way of justifying the destruction of a cation in the ISM is to invoke dissociative recombination reactions. However, dissociative recombinations of relatively large molecules typically produce fragments due to bond breaking between heavy atoms and to a much lesser extent, X-H breaks, with X being any atom heavier than deuterium. This effect has been demonstrated for several abundant molecules such as methanol. The dissociative recombination of protonated methanol was found to yield methanol + H just in a 3\% \citep{Geppert2006}. In the same way, the theoretical study of the dissociative recombination of protonated formamide \citep{ayouz2019} (NH$_{2}$CHOH$^{+}$) shows that this process only generates 15\% of formamide (NH$_{2}$CHO) + H, whilst the remaining 85\% generates HCO + NH$_{2}$ + H. Similar results were obtained in the experimental study of the dissociative recombination of cat1 and/or cat2 and cat3  (\citealt{vigren2010}; note that they do not distinguish between isomers). In this case, only 13\% of the products lead to X-H bond breaking, while the remaining 87\% of the products undergo bond breaking between heavy atoms. Note that 13\% of the H-X breaking bonds involve up to 7 exit channels, where only one of them leads to HCOOH + H.  

\citet{Herbst1985} proposed that the main route for the formation of cat1 (and/or cat2 and cat3) comes from the radiative association of HCO$^{+}$ and H$_{2}$O. However, we investigated the MEP for the association of both fragments and the barrierless reaction led to the formation of the complex CO-H$_{3}$O$^{+}$. In this process the fragment HCO$^{+}$ donates a proton to H$_{2}$O through an exothermic process that yields the dissociation of the complex giving CO and H$_{3}$O$^{+}$. This fast association-dissociation between a H$^{+}$-donor and a neutral partner should show similar rate coefficients to those calculated herein for the formation of cat1 from HCO$^{+}$ and HCOOH (Figure 2). These discrepancies invite to revisit the formation of cat1 (and/or cat2 and cat3) \citep{Herbst1985} and consequently the chemical models that include that reaction \citep{vigren2010}.

All these results point to the idea that the dissociative recombination of cat1 is not the main route for the formation of c-HCOOH and t-HCOOH. Instead, a fragmentation into smaller fragments is expected. Therefore, the most probable way of enhancing t-HCOOH with respect to c-HCOOH in dark clouds is the destruction of cat1, cat2 and cat3 after reacting with abundant molecules acting as a base such as NH$_{3}$.

Is important to highlight that our new (SAB) mechanism proposed here is not only applicable to HCOOH but to any organic molecule present in the ISM. Therefore, this mechanism could be behind the formation of high energy isomers whose provenance is unknown yet.

\begin{table*}[h!]
\setlength{\tabcolsep}{1pt} 
\caption{Fitting parameters for the destruction of cat1 by reaction with NH$_{3}$ for the temperature range indicated in parenthesis.}             
\label{tab2}      
\centering           
\begin{tabular}{c c c c }  
\hline\hline    
\multicolumn{4}{c}{cat1 + NH$_{3}$}\\
\hline
 Products & $\alpha$ (cm${^3}$ molecule$^{-1}$ s$^{-1}$) & $\beta$ & $\gamma$ (K)\\
\hline         
t-HCOOH + NH$_{4}^{+}$ (15-400 K) & 3.82x10$^{-9}$ & -0.20 & 2.23\\
c-HCOOH + NH$_{4}^{+}$ (15-400 K) & 1.26x10$^{-9}$ & -0.12 & -2.22\\
CO + NH$_{4}^{+}$ + H$_{2}$O (15-120 K) & 3.60x10$^{-18}$ & 4.32 & -155.34\\
CO + NH$_{4}^{+}$ + H$_{2}$O (130-400 K) & 4.41x10$^{-19}$ & 11.36 & -1.22x10$^{2}$\\
HCONH$_{3}^{+}$ + H$_{2}$O (15-120 K) & 1.45x10$^{-18}$ & 5.16 & 532.02\\
HCONH$_{3}^{+}$ + H$_{2}$O (130-400 K) & 3.22x10$^{-19}$ & 9.53 & -164.02\\
\hline
\end{tabular}
\end{table*}

\begin{table}[h!]
\setlength{\tabcolsep}{3.5pt} 
\caption{Product branching ratios (\%) for the destruction of cat1 by reaction with NH$_{3}$.}             
\label{tab3} 
\centering           
\begin{tabular}{c c c c c}  
\hline\hline    
T (K) & t-HCOOH & c-HCOOH & CO + NH$_{4}^{+}$  & HCONH$_{3}^{+}$ \\
 &  &  &  + H$_{2}$O & + H$_{2}$O\\
\hline         
15	&	74.28	&	25.72	&	0.00	&	0.00	\\
20	&	75.23	&	24.77	&	0.00	&	0.00	\\
25	&	75.70	&	24.30	&	0.00	&	0.00	\\
30	&	75.96	&	24.04	&	0.00	&	0.00	\\
35	&	76.11	&	23.89	&	0.00	&	0.00	\\
40	&	76.20	&	23.80	&	0.00	&	0.00	\\
45	&	76.24	&	23.76	&	0.00	&	0.00	\\
50	&	76.30	&	23.70	&	0.00	&	0.00	\\
55	&	76.27	&	23.73	&	0.00	&	0.00	\\
60	&	76.25	&	23.75	&	0.00	&	0.00	\\
65	&	76.22	&	23.78	&	0.00	&	0.00	\\
70	&	76.19	&	23.81	&	0.00	&	0.00	\\
75	&	76.16	&	23.84	&	0.00	&	0.00	\\
80	&	76.13	&	23.87	&	0.00	&	0.00	\\
85	&	76.10	&	23.90	&	0.00	&	0.00	\\
90	&	76.07	&	23.93	&	0.00	&	0.00	\\
95	&	76.04	&	23.96	&	0.00	&	0.00	\\
100	&	76.01	&	23.99	&	0.00	&	0.00	\\
110	&	75.95	&	24.05	&	0.00	&	0.00	\\
120	&	75.89	&	24.11	&	0.00	&	0.00	\\
130	&	75.83	&	24.17	&	0.00	&	0.00	\\
150	&	75.72	&	24.28	&	0.00	&	0.00	\\
170	&	75.60	&	24.40	&	0.00	&	0.00	\\
190	&	75.42	&	24.58	&	0.00	&	0.00	\\
200	&	75.38	&	24.62	&	0.00	&	0.00	\\
220	&	75.13	&	24.87	&	0.00	&	0.00	\\
240	&	75.12	&	24.88	&	0.00	&	0.00	\\
260	&	75.01	&	24.99	&	0.00	&	0.00	\\
280	&	74.94	&	25.06	&	0.00	&	0.00	\\
300	&	74.85	&	25.15	&	0.00	&	0.00	\\
320	&	74.77	&	25.23	&	0.00	&	0.00	\\
340	&	74.69	&	25.31	&	0.00	&	0.00	\\
360	&	74.61	&	25.39	&	0.00	&	0.00	\\
380	&	74.54	&	25.46	&	0.00	&	0.00	\\
400	&	74.47	&	25.53	&	0.00	&	0.00	\\
\hline
\end{tabular}
\end{table}

\section{Conclusions}

In this work we propose a novel cyclic reaction mechanism for destruction and backward formation (baptized as sequential acid-base (SAB) mechanism) to explain the occurence of c-HCOOH in dark molecular clouds. Our high-level $ab$ $initio$ quantum chemical and kinetic calculations show that the destruction of both isomers of HCOOH by reaction with HCO$^{+}$ to give the cations (cat1, cat2 and cat3), is a very efficient process. Following these reactions, the most abundant cation (cat1) reacts with NH$_{3}$ yielding c-HCOOH and t-HCOOH back with a cis/trans ratio of 25.7\%. This result explains the presence of c-HCOOH in dark molecular clouds, although this ratio is far from the 6\% ratio reported by observations. It may be that for reaching a ratio of 6\% of c-HCOOH with respect to t-HCOOH, the destruction of cat2 and cat3 should also be considered. In addition, further destruction mechanisms of the three cations with other abundant chemical species acting as bases could be considered. The detection of the cation cat1 in the ISM would reinforce our proposal and would provide valuable information for the chemistry of molecular dark clouds.

This work brings a new approach in the destruction and formation of isomers in the ISM, which involves two abundant interstellar species, HCO$^{+}$ and NH$_{3}$. This mechanistic approach can be applied to any molecule present in the gas phase of molecular clouds, and it promises to be a key process for understanding the observed ratio of the isomers of organic molecules measured in the ISM.

\begin{acknowledgements}

J.G.d.l.C. acknowledges the Spanish State Research Agency (AEI) through project number MDM-2017-0737 Unidad de Excelencia “María de Maeztu”—Centro de Astrobiología
(CSIC-INTA). J.G.d.l.C., I.J.-S., J.M.-P. and L.C. acknowledge support from grant No. PID2019-105552RB-C41 by the Spanish Ministry of Science and Innovation/State Agency of Research MCIN/AEI/10.13039/501100011033 and by “ERDF A way of making Europe”. J.C.C. acknowledges the Junta de Extremadura and European Regional Development Fund, Spain, Project No. GR21032. G. M. acknowledges the support of the Alexander von Humboldt Foundation through a postdoctoral research grant. V.M.R. has received support from the project RYC2020-029387-I funded by MCIN/AEI /10.13039/501100011033, and from the Comunidad de Madrid through the Atracci\'on de Talento Investigador Modalidad 1 (Doctores con experiencia) Grant (COOL: Cosmic Origins Of Life; 2019-T1/TIC-5379).
Computational assistance was provided by the Supercomputer facilities of LUSITANIA founded by Cénits and Computaex Foundation.

\end{acknowledgements}

\bibliographystyle{aa}
\bibliography{42287corr}

\begin{appendix}
\section{Calculation of rate constants for astrochemical models} \label{sec:appendixA}

The methodology we propose here is employed to provide a set of phenomenological rate constants to be introduced in astrochemical models. In the case of gas phase reactions, these rate constants can be obtained from experiments, calculations or from approximated theories. In the particular case of ion-molecule reactions, the Su-Chesnavich approach to compute the rate constant is employed \citep{Su1981}. Under this theory, the (global) rate constant for an ion-molecule reaction depends solely on the isotropic polarizability of the neutral molecule $\alpha$ and its dipole moment $\mu$. The Su-Chesnavich rate constant thus reads:

\begin{equation}
k_{d} = 
\begin{cases}
k_{L}\left[ \dfrac{(x+0.5090)^{2}}{10.526} + 0.9754 \right] & x < 2 \\
k_{L}\left( 0.4767x + 0.6200 \right) &  x \geq 2
\end{cases}
\end{equation}

with $k_{L}$:

\begin{equation}
  k_{L} = 2 \pi e \sqrt{\dfrac{\alpha}{M}},
\end{equation}

and $x$:

\begin{equation}
  x = \dfrac{\mu}{\sqrt{2 \alpha k_{B} T}}.
\end{equation}

with $M$ the reduced mass of the reacting fragments, $e$ the charge of the electron, $k_B$ the Boltzmann constant, and $T$ the gas temperature. The Su-Chesnavich relation has been applied to expand astrochemical networks significantly \citep{Woon2009, wakelam_reaction_2010} or in the evaluation of individual rate constants \citep{Shingledecker2020}. While the rate constants obtained for this method are approximated, they serve to shed light about destruction routes of interstellar molecules in the absence of more sophisticated theories or experiments.

To put in context the rate constants derived in this work with the ones obtained from the Su-Chesnavich approach we have calculated the Su-Chesnavich rate constants using the magnitudes obtained at our level of theory DSD-revPBEP86-D3(BJ)/aug-cc-pVTZ. The derived magnitudes for the neutral reactions of our study are collated in Table \ref{tab:polar}. The rate constants (and associated reduced mass) is considered for reactions \ce{c-HCOOH + HCO+ -> Products}, \ce{t-HCOOH + HCO+ -> Products}, and \ce{cat1 + NH3 -> Products}. We indicate products in this notation because Su-Chesnavich rate constants are phenomenological rate constants that are independent of the obtained products.

\begin{table}[h!]
\caption{Dipole moments ($\mu$ in Debye, D) and diagonal elements of the polarizability tensor ($\alpha_{ii}$ in \AA$^{3}$) for the neutral species considered in this work at the DSD-PBEP86-D3(BJ)/aug-cc-pVTZ level of theory.}             
\label{tab:polar} 
\centering    
\begin{tabular}{c c c c c }        
\hline\hline                 
Species & $\mu$ & $\alpha_{xx}$ & $\alpha_{yy}$ & $\alpha_{zz}$ \\    
\hline 
c-HCOOH & 3.90 & 4.43 & 3.32 & 2.41 \\
t-HCCOH & 1.49 & 4.06 & 3.50 & 2.39 \\
NH$_{3}$ & 1.52 & 2.03 & 2.03 & 2.27 \\
\hline                                   
\end{tabular}
\end{table}

In Figure A.1, we present the comparison between the rate constants obtained in Section \ref{results} with those calculated using the Su-Chesnavich method. In light of these results, we conclude that the Su-Chesnavich formula in some cases under- and overestimates the reaction rate constants, but provides a qualitative picture that confirms the trends obtained using more sophisticated methods. Moreover, it gives a physical explanation for the increased reactivity of c-HCOOH over t-HCOOH, finding its origin in the different dipole moment of both isomers ($\alpha$ is very similar between isomers).

\begin{figure}[h!]
  \centering
  \includegraphics[width=9.5cm,height=12cm]{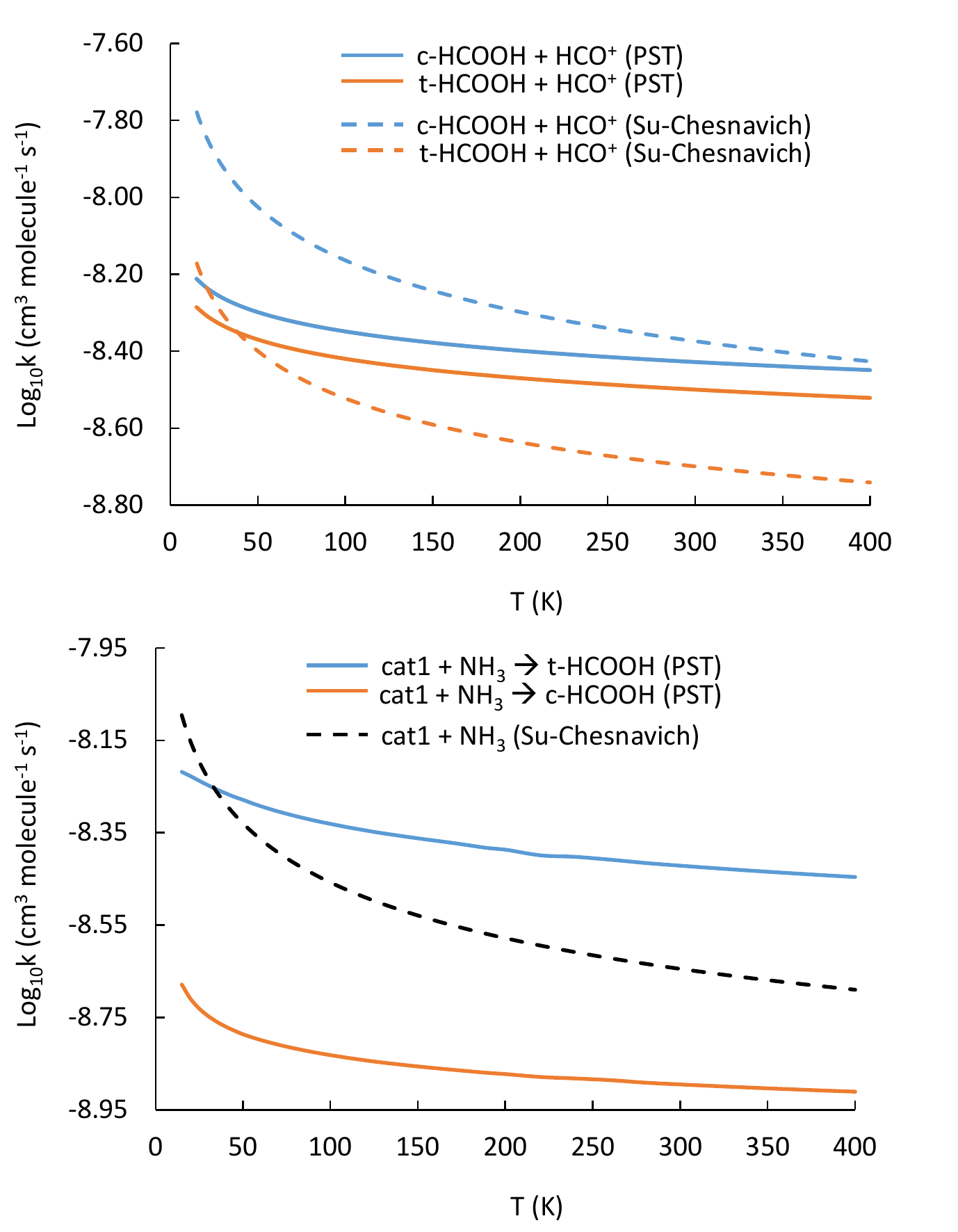}
  \caption{Comparison between the global rate constants having employed PST and those obtained with the Su-Chesnavich formula.}
\end{figure}

\section{Product branching ratios} \label{sec:appendixB}

\begin{table*}
\setlength{\tabcolsep}{3.5pt} 
\caption{Product branching ratios (\%) for the destruction of c-HCOOH by reaction with HCO$^{+}$.}             
\label{tab} 
\centering           
\begin{tabular}{c c c c c c c}  
\hline\hline    
\multicolumn{6}{c}{c-HCOOH + HCO$^{+}$}\\
\hline
T (K) & t-HCOOH & cat1 + CO & cat2 + CO & cat3 + CO & cat4 + CO & CO-H$_{3}$O$^{+}$\\
      & + CO    &  &  &  &  & + CO\\
\hline         
15	&	0.00	&	68.91	&	0.51	&	30.58	&	0.00	&	0.00	\\
20	&	0.00	&	68.91	&	0.50	&	30.60	&	0.00	&	0.00	\\
25	&	0.00	&	68.90	&	0.49	&	30.60	&	0.00	&	0.00	\\
30	&	0.00	&	68.90	&	0.49	&	30.61	&	0.00	&	0.00	\\
35	&	0.00	&	68.90	&	0.49	&	30.61	&	0.00	&	0.00	\\
40	&	0.00	&	68.90	&	0.49	&	30.61	&	0.00	&	0.00	\\
45	&	0.00	&	68.90	&	0.49	&	30.61	&	0.00	&	0.00	\\
50	&	0.00	&	68.89	&	0.49	&	30.62	&	0.00	&	0.00	\\
55	&	0.00	&	68.89	&	0.49	&	30.62	&	0.00	&	0.00	\\
60	&	0.00	&	68.89	&	0.49	&	30.61	&	0.00	&	0.00	\\
65	&	0.00	&	68.89	&	0.49	&	30.61	&	0.00	&	0.00	\\
70	&	0.00	&	68.89	&	0.49	&	30.61	&	0.00	&	0.00	\\
75	&	0.00	&	68.89	&	0.50	&	30.61	&	0.00	&	0.00	\\
80	&	0.00	&	68.89	&	0.50	&	30.61	&	0.00	&	0.00	\\
85	&	0.00	&	68.89	&	0.50	&	30.61	&	0.00	&	0.00	\\
90	&	0.00	&	68.89	&	0.50	&	30.61	&	0.00	&	0.00	\\
95	&	0.00	&	68.89	&	0.50	&	30.61	&	0.00	&	0.00	\\
100	&	0.00	&	68.89	&	0.50	&	30.61	&	0.00	&	0.00	\\
110	&	0.00	&	68.89	&	0.50	&	30.61	&	0.00	&	0.00	\\
120	&	0.00	&	68.89	&	0.51	&	30.61	&	0.00	&	0.00	\\
130	&	0.00	&	68.88	&	0.51	&	30.61	&	0.00	&	0.00	\\
150	&	0.00	&	68.88	&	0.52	&	30.60	&	0.00	&	0.00	\\
170	&	0.00	&	68.88	&	0.52	&	30.60	&	0.00	&	0.00	\\
190	&	0.00	&	68.88	&	0.53	&	30.60	&	0.00	&	0.00	\\
200	&	0.00	&	68.87	&	0.53	&	30.60	&	0.00	&	0.00	\\
220	&	0.00	&	68.87	&	0.54	&	30.59	&	0.00	&	0.00	\\
240	&	0.00	&	68.87	&	0.54	&	30.59	&	0.00	&	0.00	\\
260	&	0.00	&	68.86	&	0.55	&	30.58	&	0.00	&	0.00	\\
280	&	0.00	&	68.86	&	0.56	&	30.58	&	0.00	&	0.00	\\
300	&	0.00	&	68.85	&	0.57	&	30.58	&	0.00	&	0.00	\\
320	&	0.00	&	68.85	&	0.58	&	30.57	&	0.00	&	0.00	\\
340	&	0.00	&	68.84	&	0.59	&	30.57	&	0.00	&	0.00	\\
360	&	0.00	&	68.84	&	0.60	&	30.56	&	0.00	&	0.00	\\
380	&	0.00	&	68.83	&	0.61	&	30.56	&	0.00	&	0.00	\\
400	&	0.00	&	68.82	&	0.62	&	30.56	&	0.00	&	0.00	\\
\hline
\end{tabular}
\end{table*}

\begin{table*}
\setlength{\tabcolsep}{3.5pt} 
\caption{Product branching ratios (\%) for the destruction of t-HCOOH by reaction with HCO$^{+}$.}             
\label{tab} 
\centering           
\begin{tabular}{c c c c c c c}  
\hline\hline    
\multicolumn{6}{c}{t-HCOOH + HCO$^{+}$}\\
\hline
T (K) & cat1 + CO & cat2 + CO & cat3 + CO & cat4 + CO & CO-H$_{3}$O$^{+}$ + CO \\
 \hline         
15	&	71.90	&	0.00	&	0.36	&	17.98	&	9.76	\\
20	&	69.95	&	0.00	&	0.34	&	19.89	&	9.83	\\
25	&	68.83	&	0.00	&	0.32	&	20.98	&	9.85	\\
30	&	68.15	&	0.00	&	0.32	&	21.66	&	9.87	\\
35	&	67.70	&	0.00	&	0.32	&	22.10	&	9.89	\\
40	&	67.40	&	0.00	&	0.31	&	22.38	&	9.90	\\
45	&	67.20	&	0.00	&	0.31	&	22.57	&	9.91	\\
50	&	66.90	&	0.00	&	0.31	&	22.87	&	9.92	\\
55	&	66.91	&	0.00	&	0.31	&	22.85	&	9.93	\\
60	&	66.92	&	0.00	&	0.31	&	22.83	&	9.93	\\
65	&	66.93	&	0.00	&	0.32	&	22.81	&	9.94	\\
70	&	66.95	&	0.00	&	0.32	&	22.79	&	9.95	\\
75	&	66.96	&	0.00	&	0.32	&	22.77	&	9.95	\\
80	&	66.97	&	0.00	&	0.32	&	22.74	&	9.96	\\
85	&	66.99	&	0.00	&	0.32	&	22.72	&	9.97	\\
90	&	67.01	&	0.00	&	0.32	&	22.70	&	9.97	\\
95	&	67.02	&	0.00	&	0.32	&	22.68	&	9.98	\\
100	&	67.04	&	0.00	&	0.32	&	22.65	&	9.98	\\
110	&	67.08	&	0.00	&	0.33	&	22.60	&	9.99	\\
120	&	67.12	&	0.00	&	0.33	&	22.55	&	10.00	\\
130	&	67.17	&	0.00	&	0.33	&	22.50	&	10.01	\\
150	&	67.26	&	0.00	&	0.33	&	22.38	&	10.02	\\
170	&	67.36	&	0.00	&	0.34	&	22.26	&	10.04	\\
190	&	67.47	&	0.00	&	0.34	&	22.14	&	10.05	\\
200	&	67.53	&	0.00	&	0.35	&	22.07	&	10.05	\\
220	&	67.65	&	0.00	&	0.35	&	21.93	&	10.06	\\
240	&	67.79	&	0.00	&	0.36	&	21.78	&	10.07	\\
260	&	67.93	&	0.00	&	0.36	&	21.62	&	10.08	\\
280	&	68.09	&	0.00	&	0.37	&	21.45	&	10.08	\\
300	&	68.25	&	0.00	&	0.38	&	21.28	&	10.09	\\
320	&	68.42	&	0.00	&	0.39	&	21.09	&	10.10	\\
340	&	68.60	&	0.00	&	0.39	&	20.90	&	10.10	\\
360	&	68.79	&	0.00	&	0.40	&	20.70	&	10.10	\\
380	&	68.98	&	0.00	&	0.41	&	20.50	&	10.11	\\
400	&	69.18	&	0.00	&	0.42	&	20.29	&	10.11	\\
\hline
\end{tabular}
\end{table*}

\end{appendix}

\end{document}